\documentclass[12pt]{article}
\usepackage{graphicx}
\def\Title#1{\begin{center} {\Large #1 } \end{center}}
\def\Author#1{\begin{center}{ \sc #1} \end{center}}
\def\Address#1{\begin{center}{ \it #1} \end{center}}

\newenvironment{Abstract}{\begin{quotation}  }{\end{quotation}}
\newenvironment{Presented}{\begin{quotation} \begin{center} 
             PRESENTED AT\end{center}\bigskip 
      \begin{center}\begin{large}}{\end{large}\end{center} \end{quotation}}
\def\Acknowledgements{\bigskip  \bigskip \begin{center} \begin{large}
             \bf ACKNOWLEDGEMENTS \end{large}\end{center}}

\begin{document}
\begin{titlepage}


\Title{Experimental determination of $V_{us}$ from kaon decays}

\vfill

\Author{Matthew Moulson}
\Address{Laboratori Nazionali di Frascati dell'INFN\\
Via E. Fermi, 40, 00044 Frascati RM, Italy}

\vfill

\begin{Abstract}
During the last few years, new experimental and theoretical results have 
allowed ever-more-stringent tests of the Standard Model to be performed 
using kaon decays. This overview of recent progress includes updated results 
for the evaluation of the CKM matrix element $V_{us}$ and related tests of 
CKM unitarity from experimental data on kaon decays.
\end{Abstract}

\vfill

\begin{Presented}
8th International Workshop on the CKM Unitarity Triangle (CKM 2014), 
Vienna, Austria, September 8-12, 2014
\end{Presented}

\vfill

\end{titlepage}

\def\thefootnote{\fnsymbol{footnote}}
\setcounter{footnote}{0}

\section{Introduction}

If indeed the $W$ couples to quarks and leptons via a single, 
universal gauge coupling, then for universality to be observed as the 
equivalence of the Fermi constant $G_F$ as measured in muon and hadron decays, 
the CKM matrix must be unitary. Currently, the most stringent test of 
CKM unitarity is obtained from the first-row relation
$V_{ud}^2 + V_{us}^2 + V_{ub}^2 \equiv 1 + \Delta_{\rm CKM}$.
During the period spanning 2003 to 2010, a wealth of new measurements
of $K_{\ell3}$ and $K_{\ell2}$ decays, together with
steady theoretical progress, made possible precision tests of the Standard 
Model (SM) based on this relation. In a 2010 evaluation of $V_{us}$, the 
FlaviaNet Working Group on Kaon Decays set bounds on $\Delta_{\rm CKM}$
at the level of 0.1\%~\cite{FlaviaNet+10:Vus}, which translate into bounds 
on the effective scale of new physics on the order of 10~TeV~\cite{CJGA10:eff}.
Since 2010, there have been a few significant new measurements; the purpose 
of the present review is to update the FlaviaNet evaluation by including 
them. During the last four years, there have also been important theoretical
developments. Advances in algorithmic sophistication and computing power
have lead to a number of new, high-precision lattice QCD estimates of the 
hadronic constants $f_+(0)$ and $f_K/f_\pi$, which enter into the 
determination of $V_{us}$ from $K_{\ell 3}$ and $K_{\mu2}$  decays,
respectively.  The Flavor Lattice Averaging Group (FLAG), whose members
include representatives from the major lattice collaborations, provides 
synthesis of the results of lattice calculations in the form of a biannual 
review. The most recent edition \cite{FLAG+14:review} is a useful source for the
recommended values of the lattice constants entering into the evaluation 
of $V_{us}$. Theoretical issues and updates since the FLAG review were
also summarized at this conference \cite{Gar14:CKM}.

\section{New experimental inputs}
\label{sec:expt}

\subsection{Branching ratios and lifetimes}

The experimental inputs for the determination of $V_{us}$ from $K_{\ell3}$
decays are the rates and form factors for the decays of both charged and 
neutral kaons. Existing data allow $V_{us}$ to be evaluated from the rates
for the $K_{e3}$ decays of the $K_S$, $K_L$, and $K^\pm$, and for the $K_{\mu3}$ 
decays of the $K_L$ and $K^+$. Because the best values for the 
leptonic and semileptonic decay rates are obtained from fits to the 
branching ratio (BR) and lifetime measurements for all decay modes with 
appreciable BRs, any new BR measurements are potentially interesting 
inputs to the analysis.

Since the 2010 review, there have been two new measurements of the $K_S$ 
lifetime, $\tau_{K_S}$: one from KLOE, from the vertex
distribution for $K_S\to\pi^+\pi^-$ decays in $e^+e^-\to\phi\to K_SK_L$
events \cite{KLOE+11:KSlife}, and one from KTeV, from a comprehensive 
reanalysis of the experiment's 
data set to obtain a new value for $\mathrm{Re}\,\varepsilon'/\varepsilon$
\cite{KTeV+11:epspr}.
The KTeV analysis can be performed with or without assuming 
$CPT$ symmetry; the present update makes use of the results 
obtained without the $CPT$ assumption. 
The largest effect of these updates on the $K_S$ fit is to reduce the
uncertainty on $\tau_{K_S}$, the value of which changes from $89.59(6)$~ps
\cite{FlaviaNet+10:Vus} to $89.58(4)$~ps.
The result for ${\rm BR}(K_{S\,e3})$ from the $K_S$ fit is essentially
determined by the KLOE result for 
${\rm BR}(K_{S\,e3})/{\rm BR}(K_S\to\pi^+\pi^-)$
\cite{KLOE+06:KSe3}.

There have been no new $K_L$ branching ratio measurements 
since the 2010 review, but, as noted above, KTeV has a new value for
$\mathrm{Re}\,\varepsilon'/\varepsilon$. For the present analysis,
this result is averaged with the 2002 result from NA48 \cite{NA48+02:Reps01}
to provide an effective measurement of 
$\mathrm{BR}(K_L\to\pi^0\pi^0)/\mathrm{BR}(K_L\to\pi^+\pi^-)$
obtained without assuming $CPT$ symmetry. This is now used in the $K_L$ 
fit in place of the PDG ETAFIT result. The $K_L$ fit results from the 
2010 review are essentially unchanged. The consistency of the input data
is generally poor---the fit gives $\chi^2/{\rm ndf} = 19.8/12$ ($P=7.0\%$).

\begin{table}
\begin{center}
\begin{tabular}{lcc}
\hline\hline
Parameter & Value & $S$ \\
\hline
${\rm BR}(\mu\nu)$        & 63.58(11)\%   & 1.1 \\
${\rm BR}(\pi\pi^0)$      & 20.64(7)\%    & 1.1 \\
${\rm BR}(\pi\pi^+\pi^-)$ & 5.56(4)\%     & 1.0 \\
${\rm BR}(\pi^0 e\nu)$    & 5.088(27)\%   & 1.2 \\ 
${\rm BR}(\pi^0 \mu\nu)$  & 3.366(30)\%   & 1.9 \\
${\rm BR}(\pi\pi^0\pi^0)$ & 1.764(25)\%   & 1.0 \\
$\tau_{K^\pm}$            & 12.384(15)~ns & 1.2 \\ 
\hline\hline
\end{tabular}
\end{center}
\caption{\label{tab:kpm}
Complete results of updated fit to $K^\pm$ BR and lifetime measurements.
Scale factors are calculated using the PDG prescription \cite{PDG+14:RPP}.}
\end{table}
A new measurement of ${\rm BR}(K^\pm\to\pi^\pm\pi^+\pi^-)$ from 
KLOE-2~\cite{KLOE2+14:Kppp} fills a significant gap in the $K^\pm$ data set.
For the 2010 review, the only constraint in the $K^\pm$ fit on
the value of ${\rm BR}(K^\pm\to\pi^\pm\pi^+\pi^-)$ was from a 1965 
bubble-chamber measurement with a 3\% uncertainty \cite{B+65:Kppp} 
that was slated for removal upon the availability
of a new measurement with a thorough evaluation of systematic errors.
The $K^\pm\to\pi^\pm\pi^+\pi^-$ decay is not easy to reconstruct at 
KLOE because of the low momentum of the pion tracks. Nevertheless, 
KLOE-2 obtains a result with an overall precision of 0.7\%: 
${\rm BR}(K^\pm\to\pi^\pm\pi^+\pi^-) = 
0.005565(31)_{\rm stat}(25)_{\rm sys}$. This result is claimed to be 
fully inclusive of radiation,
but the radiative corrections are handled somewhat differently than
in earlier KLOE analyses. The inclusion of this measurement
in place of the measurement from \cite{B+65:Kppp} 
significantly reduces the uncertainty on the fit result for 
${\rm BR}(K^\pm_{\mu2})$, 
as suggested by the large correlation ($\rho=-0.75$) in the previous 
fit between the values for ${\rm BR}(K^\pm_{\mu2})$ and 
${\rm BR}(K^\pm\to\pi^\pm\pi^+\pi^-)$.

The updated fit to 6 BRs 
and the $K^\pm$ lifetime uses 17 input measurements, has 1 constraint, 
and gives $\chi^2/{\rm ndf} = 25.5/11$ ($P=0.78\%$). 
The full results of the fit are listed in Table~\ref{tab:kpm}.
The poor $\chi^2$ 
value is largely the result of the large residuals for the KLOE BR 
measurements for $K^\pm_{e3}$ and $K^\pm_{\mu3}$
($-2.3\sigma$ and $-3.4\sigma$, respectively). 
For comparison, the current PDG fit \cite{PDG+14:RPP} to 7 BRs and 
the $K^\pm$ lifetime uses several of the older measurements not used 
in the present analysis (32 measurements in all), and gives
$\chi^2/{\rm ndf} = 51.8/25$ ($P=0.13\%$). 

Recently, the ISTRA+ experiment at Protvino has published a result 
for the ratio ${\rm BR}(K^-_{e3})/{\rm BR}(\pi\pi^0)$ from the analysis 
of data taken in 2001 \cite{ISTRA+14:Ke3FF}.
The sample of $K_{e3}$ events is claimed to be inclusive of inner 
bremsstrahlung (IB). However, radiative events are not generated in
the simulation used to calculate the acceptance, so there is some
uncertainty as to how inclusive the measurement is. 
In part, this is taken into account in the evaluation of the 
systematic uncertainties via variation of the angular cut used to
accept a possible IB photon. This measurement is not yet included
in the evaluation of $V_{us}\,f_+(0)$. The effect of including
it in the fit to $K^\pm$ rate data is to change the result for
${\rm BR}(K_{e3})$ from 5.088(27)\% to 5.083(27)\%,
with a similar effect on ${\rm BR}(K_{\mu3})$, and negligible
effects on the other BRs and on the fit quality.

\subsection{Form factor parameters}
\label{sec:ff}

\begin{figure}
\begin{center}
\includegraphics[height=0.35\textheight]{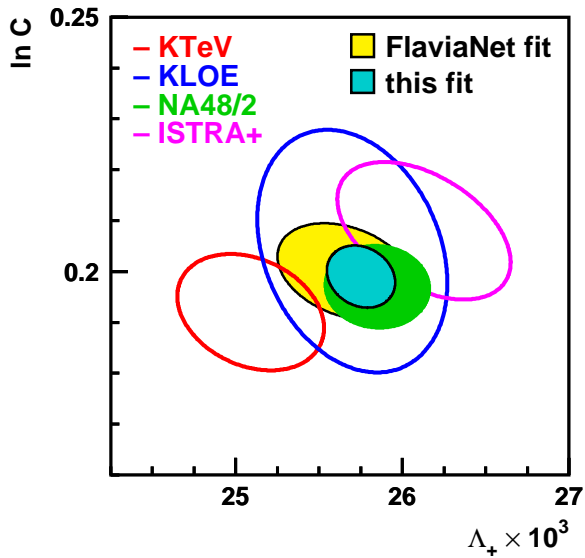}
\caption{$1\sigma$ confidence contours for form factor parameters 
($K_{e3}$-$K_{\mu3}$ averages) from dispersive fits, for different 
experiments. 
The NA48/2 result was converted by the author from the experiment's polynomial 
fit results. The FlaviaNet 2010 average and the new average, with the NA48/2
result included, are also shown.}
\label{fig:ffcomp}
\end{center}
\end{figure} 
In 2012, the NA48/2 experiment released preliminary results for the 
form factors for both $K^\pm_{e3}$ and $K^\pm_{\mu3}$ decays~\cite{Wan12:HQL}.
The fits were performed using one of two parameterizations,
\begin{itemize}
\item polynomial: $(\lambda'_+, \lambda''_+)$ for $K_{e3}$, 
$(\lambda'_+, \lambda''_+, \lambda_0)$ for $K_{\mu3}$, or
\item polar: $M_V$ for $K_{e3}$, $(M_V, M_S)$ for $K_{\mu3}$.
\end{itemize}
The FlaviaNet Working Group uses the dispersive parameterizations 
of~\cite{B+09:disp} because of the advantages described 
in~\cite{FlaviaNet+10:Vus}. NA48/2 has not performed dispersive fits.
However, it is possible to obtain approximately equivalent values
of $\Lambda_+$ and $\mathrm{ln}\,C$ from a fit to the NA48/2 measurements of
$(\lambda'_+, \lambda''_+, \lambda_0)$ using the expressions in the appendix 
of~\cite{B+09:disp} and the observation that 
$\lambda_0 \approx \lambda_0' + 3.5\lambda_0''$ \cite{KLOE+07:Km3FF}.
This is important because it helps to resolve a controversy: the older
measurements of the $K_{\mu3}$ form factors for $K_L$ decays from 
NA48~\cite{NA48+07:m3FF} are in such strong disagreement with the other 
existing measurements that they were excluded from the FlaviaNet 
averages~\cite{FlaviaNet+10:Vus}. 
The new NA48/2 measurements, on the other hand, are in good agreement with 
other measurements, as seen from Fig.~\ref{fig:ffcomp}. Including the new 
NA48/2 results, appropriately converted for the current purposes, the 
dispersive average becomes $\Lambda_+ = (25.75\pm0.36)\times10^{-3}$, 
$\mathrm{ln}\,C = 0.1985(70)$, with $\rho = -0.202$ and $P(\chi^2) = 55\%$.
The central values of the phase space integrals barely change with this 
inclusion; the uncertainties are reduced by 30\%. 

The OKA experiment at Protvino is the successor to ISTRA+.
Recently, OKA presented preliminary results on the $K_{e3}$
form factor parameters from polynomial, pole, and dispersive fits, obtained
with a sample of 6M $K^+$ decays \cite{Obr14:ICHEP}. The results are significantly
more precise than the ISTRA+ results \cite{ISTRA+04:e3FF} and are in 
agreement with ISTRA+ and the other experiments. The OKA results will be
included in the averages for the form factor parameters in a future update,
once the systematic uncertainties have been evaluated.

\section{Evaluation of $V_{us}$ and related tests}

\begin{table}
\begin{center}
\begin{tabular}{lcccccc}
\hline\hline
              &              &         & \multicolumn{4}{@{}c@{}}{\small Approx contrib to \% err} \\
Mode          & $V_{us}\,f_+(0)$ & \% err  & BR   & $\tau$ & $\Delta$ & $I$  \\ 
\hline
$K_{L\,e3}$    & 0.2163(6)  & 0.26    & 0.09 & 0.20   & 0.11     & 0.05 \\ 
$K_{L\,\mu3}$  & 0.2166(6)  & 0.28    & 0.15 & 0.18   & 0.11     & 0.06 \\
$K_{S\,e3}$    & 0.2155(13) & 0.61    & 0.60 & 0.02   & 0.11     & 0.05 \\
$K^\pm_{e3}$   & 0.2172(8)  & 0.36    & 0.27 & 0.06   & 0.23     & 0.05 \\
$K^\pm_{\mu3}$ & 0.2170(11) & 0.51    & 0.45 & 0.06   & 0.23     & 0.06 \\
\hline\hline
\end{tabular}
\end{center}
\caption{\label{tab:Vusf}
Values of $V_{us}\,f_+(0)$ from data for different decay modes, 
with breakdown of uncertainty from different sources: branching ratio 
measurements (BR), lifetime measurements ($\tau$), long distance radiative
and isospin-breaking corrections ($\Delta$), and phase space integrals
from form factor parameters ($I$).}
\end{table}
The evaluations of $V_{us}\,f_+(0)$ for each of the five decay modes
($K_{L\,e3}$, $K_{L\,\mu3}$, $K_{S\,e3}$, $K^\pm_{e3}$, and $K^\pm_{\mu3}$)
are presented in Table~\ref{tab:Vusf}, with a breakdown of the 
uncertainties from different sources in each case.
The updated five-channel average is $V_{us}\,f_+(0) = 0.2165(4)$ with  
$\chi^2/{\rm ndf} = 1.61/4$ $(P = 81\%)$. 

In the 2010 review, $V_{us}\,f_+(0)$ was 0.2163(5), so
the fractional uncertainty has been reduced from 0.23\% to 0.18\%.
Careful comparison of Table~\ref{tab:Vusf} with the corresponding table 
in \cite{FlaviaNet+10:Vus}, however, reveals that all of the efforts
described in Section~\ref{sec:expt} contribute only marginally to this
progress. The modes for which the evaluation of $V_{us}\,f_+(0)$ is 
most improved are $K^\pm_{e3}$ ($0.2160(11)\to0.2172(8)$) and
$K^\pm_{\mu3}$ ($0.2158(14)\to0.2170(11)$). The $\sim$1$\sigma$ shift 
and reduction of the uncertainty in both cases is due to a new evaluation
of the correction for strong isospin breaking, $\Delta_{SU(2)}$.
The correction used in this analysis is
$\Delta_{SU(2)} = (2.43\pm0.20)\%$, which was calculated using the
$N_f = 2+1$ FLAG averages for the quark-mass ratios $Q = 22.6(7)(6)$
and $m_s/\hat{m} = 27.46(15)(41)$ with the isospin-limit meson masses
$M_K = 494.2$ MeV and $M_\pi = 134.8$ MeV \cite{FLAG+14:review} 
(see \cite{C+12:KaonRev} for discussion and notation). The 2010 review
used the result from the chiral perturbation theory analysis of information 
on light quark masses, $\Delta_{SU(2)} = (2.9\pm0.4)\%$ \cite{KN08:Kl3FF};
the uncertainty on this value was a leading contribution to the 
uncertainty on $V_{us}\,f_+(0)$ from the charged-kaon modes.
The strong isospin-breaking correction is still a significant source 
of uncertainty for these modes; the value of $\Delta_{SU(2)}$ used here 
should be confirmed and improved upon. 
Perfect equality of the uncorrected results for $V_{us}\,f_+(0)$ 
from charged and neutral modes would require $\Delta_{SU(2)} = 2.83(38)\%$.

A value for $f_+(0)$ is needed to obtain the value of $V_{us}$. 
As lattice calculations improve to include more dynamical flavors, 
smaller pion masses, and better systematics, there is some suggestion 
that the results for $f_+(0)$ are increasing (in particular, see Fig.~4 
of \cite{FLAG+14:review}). For this review, separate reference values are used
for $N_f = 2+1$ and $N_f = 2+1+1$. For $N_f = 2+1$, the uncorrelated FLAG  
average of the results from the 
RBC/UKQCD \cite{UKRBC+13:f0} and Fermilab/MILC \cite{MILF+13:f0}
collaborations is $f_+(0) = 0.9661(32)$. For $N_f = 2+1+1$ the
recommended FLAG value is a recently published result from
Fermilab/MILC: $f_+(0) = 0.9704(32)$ \cite{MILF+14:f0}.
Both values are significantly higher and more precise than
the reference value used in the 2010 review: $f_+(0) = 0.959(5)$.

\begin{table}
\begin{center}
\begin{tabular}{ccccc}
\hline\hline
\multicolumn{2}{@{}c@{}}{Choice of $f_+(0)$} & $V_{us}$ &
\multicolumn{2}{@{}c@{}}{$\Delta_{\rm CKM}$} \\
\hline
$N_f = 2+1$   & 0.9661(32) & 0.2241(9) & $-0.0008(6)$ & $-1.4\sigma$ \\
$N_f = 2+1+1$ & 0.9704(32) & 0.2232(9) & $-0.0012(6)$ & $-2.1\sigma$ \\
\hline\hline
\end{tabular}
\end{center}
\caption{\label{tab:Vus}
Results for $V_{us}$ and first-row unitarity test from $K_{\ell3}$ decays.}
\end{table}
The test of CKM unitarity requires a value for $|V_{ud}|$.
The survey of experimental data on $0^+\to0^+$ $\beta$ decays
from Hardy and Towner \cite{HT09:VudSFT} has been recently updated
with 24 new measurements; after a critical review of inner-bremsstrahlung 
correction schemes \cite{TH10:IBCVC}, some results have also been rejected.
The new survey gives $V_{ud} = 0.97417(21)$ \cite{HT14:SFTrev}.
The results for $V_{us}$ from $K_{\ell3}$ decays with $N_f=2+1$ and 
$N_f=2+1+1$ lattice values for $f_+(0)$ and the latest value of 
$V_{ud}$ are all listed in Table~\ref{tab:Vus}. Because of the increase 
in the results for $f_+(0)$, the previously excellent agreement with 
the expectation from first-row unitarity is no longer observed: 
$\Delta_{\rm CKM}$ is different
from zero by $-1.4\sigma$ and $-2.1\sigma$ when the $N_f=2+1$ and  
$N_f=2+1+1$ results for $f_+(0)$ are used, respectively.

Up to kinematic factors and long-distance electromagnetic corrections, 
the ratio of the (inner-bremsstrahlung inclusive) rates for $K_{\mu2}$ 
and $\pi_{\mu2}$ decays provides access to the quantity 
$V_{us}/V_{ud}\times f_{K^\pm}/f_{\pi^\pm}$.
Note that the quantity $f_{K^\pm}/f_{\pi^\pm}$ includes the effects 
of strong isospin breaking. Some lattice collaborations
make an appropriate choice of the physical point for the chiral extrapolation
and quote $f_{K^\pm}/f_{\pi^\pm}$ directly, while others quote $f_K/f_\pi$,
i.e., the ratio in the $SU(2)$ limit. In the latter cases, FLAG now 
applies corrections for strong isospin breaking
to each result to produce the averages for $f_{K^\pm}/f_{\pi^\pm}$.
This is a substantial improvement with respect to the situation in 2010,
at which time the effects of strong isospin breaking were not yet taken 
into account.\footnote{
Strong isospin breaking was taken into account for the CKM 2012 update
\cite{Mou13:CKM} 
of the FlaviaNet 2010 review. The available lattice averages were for 
$f_K/f_\pi$; corrections from \cite{CN11:Kl2IB} were made to the ratio of 
$K_{\mu2}$ to $\pi_{\mu2}$ rates for both long-distance electromagnetic 
effects and strong isospin breaking; and the experimental quantity quoted 
was $V_{us}/V_{ud}\times f_K/f_\pi$. In contrast, for the present 
determination of $V_{us}/V_{ud}\times f_{K^\pm}/f_{\pi^\pm}$,
only the electromagnetic correction of \cite{CN11:Kl2IB} is applied.}

\begin{table}
\begin{center}
\begin{tabular}{ccccc}
\hline\hline
\multicolumn{2}{@{}c@{}}{Choice of $f_{K^\pm}/f_{\pi^\pm}$} & $V_{us}/V_{ud}$ &
\multicolumn{2}{@{}c@{}}{$\Delta_{\rm CKM}$} \\
\hline
$N_f = 2+1$   & 1.192(5)   & 0.2315(10) & $-0.0001(6)$ & $-0.2\sigma$ \\
$N_f = 2+1+1$ & 1.1960(25) & 0.2308(6) & $-0.0004(5)$ & $-0.9\sigma$ \\
\hline\hline
\end{tabular}
\end{center}
\caption{\label{tab:Vusd}
Results for $V_{us}/V_{ud}$ and first-row unitarity test from 
$K_{\mu2}$ decays.}
\end{table}
Compared to the 2010 value, the value for ${\rm BR}(K^\pm_{\mu2})$ in 
Table~\ref{tab:kpm} has slightly increased and its uncertainty has 
been reduced from 0.3\% to 0.2\%, leading to the new result 
$V_{us}/V_{ud}\times f_{K^\pm}/f_{\pi^\pm} = 0.2760(4)$.
For the lattice value of $f_{K^\pm}/f_{\pi^\pm}$, once again, a distinction 
is made between results for $N_f = 2+1$ and $N_f = 2+1+1$.
The $N_f = 2+1$ FLAG average of four complete and published 
determinations of $f_{K^\pm}/f_{\pi^\pm}$ is 1.192(5) \cite{FLAG+14:review}.
For $N_f = 2+1+1$, there are two results entering into the FLAG value: from
MILC \cite{MILC+13:fKpi} and HPQCD \cite{HPQCD+13:fKpi}. For their result,
MILC generated staggered-quark ensembles with quark masses down to physical 
values. The HPQCD result is based on an alternative analysis of the MILC 
ensembles with independent methodology, leading to much smaller uncertainties. 
For $N_f = 2+1+1$, FLAG recommends the value $f_{K^\pm}/f_{\pi^\pm} = 1.194(5)$,
which covers both results; the uncertainty quoted is similar to that for the 
MILC result. After the cutoff for the FLAG review, the Fermilab/MILC 
collaboration published a study \cite{MILF+14:fKpi} updating the MILC 
result of \cite{MILC+13:fKpi}, using many of the same ensembles. This new 
result has an error budget similar to that for the HPQCD result.
It is not straightforward to combine the results for $N_f = 2+1+1$. 
For the purposes of this review,
$f_{K^\pm}/f_{\pi^\pm} = 1.1960(25)$ is used, where this
representative value is essentially the symmetrization of the result
from \cite{MILF+14:fKpi}.
The results for $V_{us}/V_{ud}$ and the first-row unitarity test 
from $K^\pm_{\mu2}$ decays, obtained with the $N_f = 2+1$ and $N_f = 2+1+1$
lattice values for $f_{K^\pm}/f_{\pi^\pm}$ and $V_{ud}$ from
\cite{HT14:SFTrev}, are presented in Table~\ref{tab:Vusd}.

The $K^\pm_{\mu2}$ data and lattice results for $f_{K^\pm}/f_{\pi^\pm}$
show better agreement with first-row unitarity than do the $K_{\ell3}$ 
data and lattice results for $f_+(0)$.
It is difficult to ascribe any significance to this difference at present,
but assuming for a moment that it points to a problem in the 
data set or the lattice calculations for $K_{\ell3}$, it would be useful 
to have additional tests to help identify the problem. Once such test 
might be provided by the Callan-Treiman relation, which links the 
value of the scalar form factor at the unphysical point 
$t_{\rm CT} = m_K^2-m_\pi^2$ to the ratio of $f_K/f_\pi$ to 
$f_+(0)$ \cite{CT66:rel,GL85:f0}.
Since $f_0(t_{\rm CT})$ is given by the form factor parameter $C$
(see Section~\ref{sec:ff}), the test provides a way to gauge the 
consistency of the value of $f_+(0)$ with the $K_{\mu3}$ slope
measurements, assuming the validity of the value for
$f_K/f_\pi$.\footnote{Here, the $N_f = 2+1+1$ value 
for $f_{K^\pm}/f_{pi^\pm}$ is 
corrected to $f_K/f_\pi = 1.1986(25)$ in the isospin limit.}
The test gives $f_+(0) = 0.980(10)$. Comparison with the lattice values
of $f_+(0)$ in Table~\ref{tab:Vus} suggests that higher values of 
$f_+(0)$ are not inconsistent with the $K_{\mu3}$ form factor parameters.
This test does not use the experimental information on the $K_{\ell3}$ rates.

\begin{figure}
\begin{center}
\includegraphics[width=0.45\textwidth]{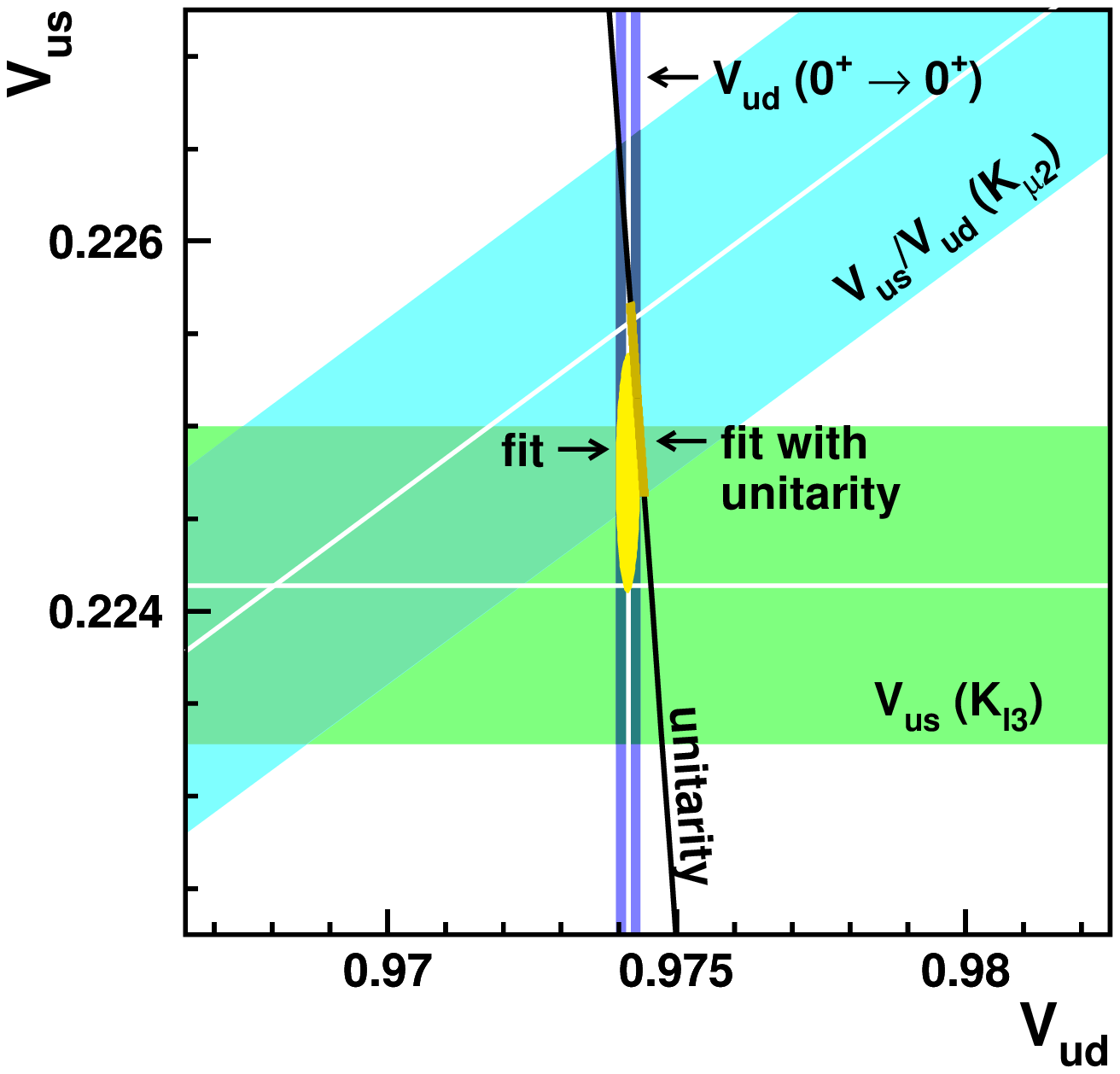}
\hfill
\includegraphics[width=0.45\textwidth]{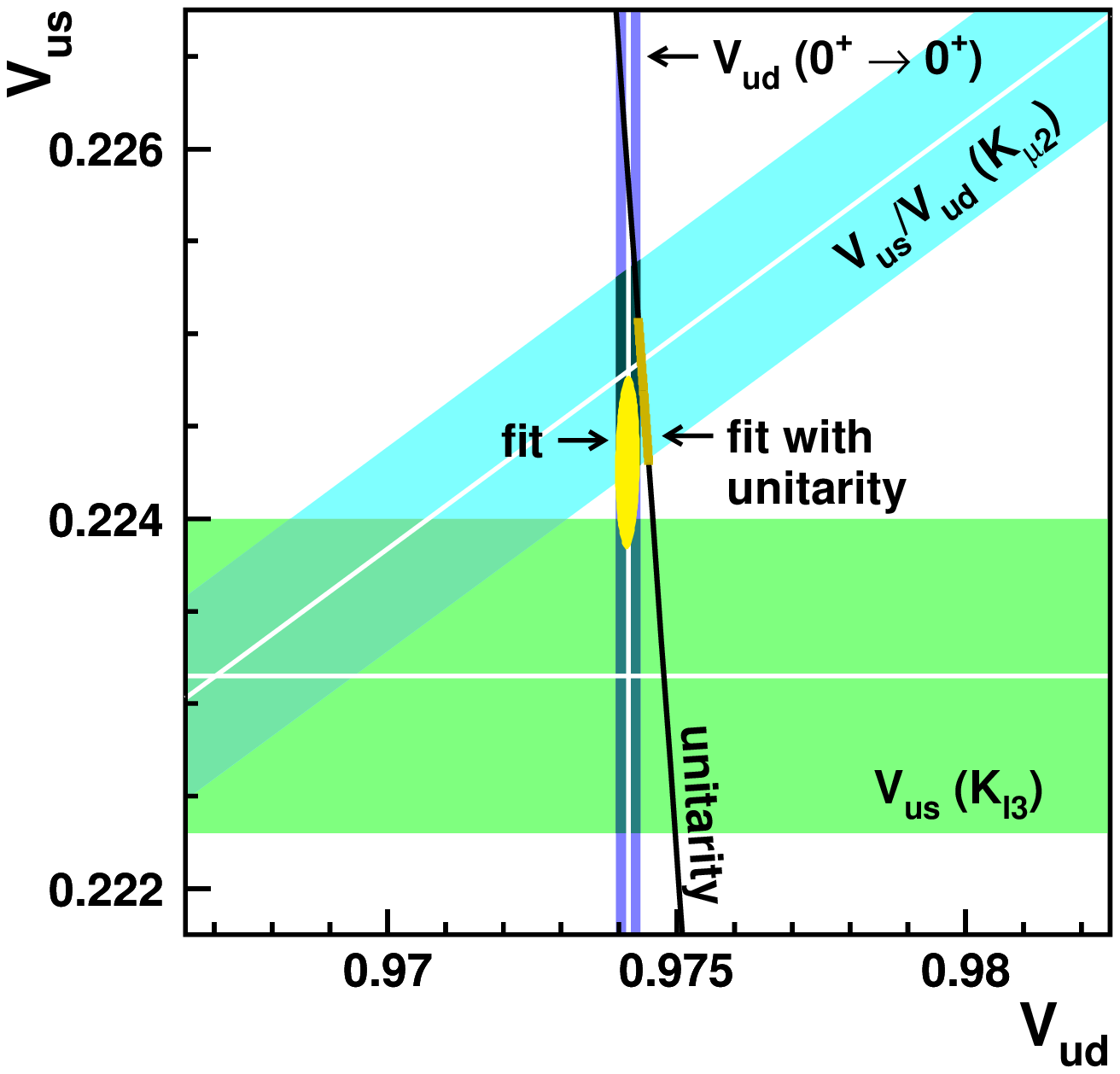}
\caption{Fits to $V_{ud}$ from $0^+\to0^+$ $\beta$ decays, $V_{us}$ from
$K_{\ell3}$ decays, and $V_{us}/V_{ud}$ from $K_{\mu2}$ decays,
with lattice values of $f_+(0)$ and $f_{K^\pm}/f_{\pi^\pm}$ from 
$N_f=2+1$ (left) and $N_f=2+1+1$ (right). The yellow 
ellipses indicate the $1\sigma$ confidence intervals in the plane of
($V_{ud}$, $V_{us}$) for the fits with with no constraints. The yellow 
line segments indicate the results obtained with the constraint 
$\Delta_{\rm CKM} = 0$.}
\label{fig:univers}
\end{center}
\end{figure} 
\begin{table}
\begin{center}
\begin{tabular}{ccc}
\hline\hline
 & $N_f = 2+1$ & $N_f = 2+1+1$ \\ 
\hline
$V_{ud}$ & 0.97416(21) & 0.97415(21) \\
$V_{us}$ & 0.2248(7) & 0.2243(5) \\
$\chi^2/{\rm ndf}$ & 1.16/1 (28.1\%) & 2.64/1 (10.4\%) \\
$\Delta_{\rm CKM}$ & $-0.0005(5)$ $(-1.0\sigma)$ & $-0.0007(5)$ $(-1.5\sigma)$
\\
\hline\hline
\end{tabular}
\end{center}
\caption{\label{tab:univers}
Results of fits to $V_{ud}$ from $0^+\to0^+$ $\beta$ decays, $V_{us}$ from
$K_{\ell3}$ decays, and $V_{us}/V_{ud}$ from $K_{\mu2}$ decays shown 
in Figure~\ref{fig:univers}.}
\end{table}
The values of $V_{ud}$ from $0^+\to0^+$ $\beta$ decays, $V_{us}$ from 
$K_{\ell3}$ decays, and $V_{us}/V_{ud}$ from $K_{\mu2}$ decays can be 
combined in a single fit to increase the sensitivity of the unitarity test, as 
illustrated in Fig.~\ref{fig:univers}. The fit can be performed with or without
the unitarity constraint, $\Delta_{\rm CKM} = 0$. 
The unconstrained fits give the results listed in Table~\ref{tab:univers}.
These fits give $\Delta_{\rm CKM} = -0.0005(5)$ $(-1.0\sigma)$ for the analysis 
using $N_f = 2+1$ lattice results, and $\Delta_{\rm CKM} = -0.0007(5)$ 
$(-1.5\sigma)$ for the analysis using $N_f = 2+1+1$ results. The corresponding
result in 2010 (for $N_f=2+1$) was $V_{us} = 0.2253(9)$ and
$\Delta_{\rm CKM} = -0.0001(6)$.
The decreased consistency with the first-row unitarity hypothesis is 
almost entirely due to the changes in the lattice results for 
$f_+(0)$. All new experimental results combined have a 
marginal effect on the results of the unitarity test, and good agreement
with unitarity is still observed for $K_{\mu2}$ decays.

While it would be premature to claim that the 
unitarity residuals for $K_{\ell3}$ are significant, the question arises: 
are hidden systematics in the data and calculations becoming important 
as the stated uncertainties shrink? If so, are they mainly in the data
or the lattice calculations? Technical and technological advances in 
lattice computing have made it possible to evaluate systematics with
increasing thoroughness, and the Callan-Treiman test favors larger values
of $f_+(0)$. Meanwhile, the consistency of the fits to $K_{\ell3}$ 
rate data is creaky. Yet, the errors on the BRs from these fits are scaled
to reflect internal inconsistencies, and after this procedure, the
values of $V_{us}\,f_+(0)$ from $K_L$, $K_S$, and $K^\pm$ modes show
good agreement. There is
also a fair amount of redundancy in the $K_{\ell3}$ data set, and adding or
eliminating single measurements doesn't change the results for 
$V_{us}\,f_+(0)$ by much.

At present, $V_{us}$ is known to nearly 0.2\%. The dominating uncertainties 
are from the lattice results rather than experiment, but not by much.
Within the next decade, the uncertainties on the lattice averages for 
$f_+(0)$ and $f_K/f_\pi$ will be decreased, perhaps down to the 0.1\% 
level \cite{Tar14:LTS1}.
If this can be achieved, there will be a demand for BR and lifetime 
measurements for $K_L$, $K_S$, and $K^\pm$ with matching precision.
The experimental parameters that currently contribute more than $0.1\%$
to the uncertainty on $V_{us}$ evaluated for a single mode are the 
$K_L$ lifetime and the BRs for $K_{S\,e3}$ and $K^\pm_{\ell3}$ decays.
In order to be useful, better ${\rm BR}(K^\pm_{\ell3})$ measurements 
would also require a more precise evaluation of $\Delta_{SU(2)}$.

\Acknowledgements
The author would like to warmly thank P.~de~Simone, V.~Duk, E.~G\'amiz, 
J.~Hardy, and I.~Towner for useful discussions and comments on this
manuscript.

\end{document}